\documentclass[aip,apl,reprint,superscriptaddress,amsmath,amssymb]{revtex4-1}

\usepackage{graphicx,bm,hyperref,natbib,color}

\newcommand{\iloop}{i_{\text{loop}}}
\newcommand{\Lknano}{L_{k_{\text{nano}}}}
\newcommand{\Lgnano}{L_{g_{\text{nano}}}}
\newcommand{\Lkmicro}{L_{k_{\text{micro}}}}
\newcommand{\Lgmicro}{L_{g_{\text{micro}}}}
\newcommand{\LTloop}{L_{T_{\text{loop}}}}
\newcommand{\textBcoil}{$B_{\text{coil}}$}
\newcommand{\textBzero}{$B_{\text{0}}$}
\newcommand{\mathBcoil}{B_{\text{coil}}}
\newcommand{\mathBzero}{B_{\text{0}}}
\newcommand{\Cc}{$C_C$}
\newcommand{\Lr}{$L_R$}
\newcommand{\Cr}{$C_R$}
\newcommand{\Tc}{T_{\text{C}}}
\newcommand{\lnw}{\ell_{\text{nw}}}

\newcommand{\figref}[1]{Fig.\ \ref{#1}}
\newcommand{\myeqref}[1]{Eq.\ \ref{#1}}
\newcommand{\textmicron}{$\mu$m}
\newcommand{\mathmicron}{\mu\text{m}}
\newcommand{\papertitle}{SKIFFS: Superconducting Kinetic Inductance Field-Frequency Sensors for Sensitive
Magnetometry in Moderate Background Magnetic Fields}

\begin{document}
\title{\papertitle}
\author{A. T. Asfaw}
\email{asfaw@princeton.edu}
\affiliation{
    Department of Electrical Engineering, Princeton University, Princeton, New Jersey 08544, USA
}
\author{E. I. Kleinbaum}
\affiliation{
    Department of Electrical Engineering, Princeton University, Princeton, New Jersey 08544, USA
}
\author{T. M. Hazard}
\affiliation{
    Department of Electrical Engineering, Princeton University, Princeton, New Jersey 08544, USA
}
\author{A. Gyenis}
\affiliation{
    Department of Electrical Engineering, Princeton University, Princeton, New Jersey 08544, USA
}
\author{A. A. Houck}
\affiliation{
    Department of Electrical Engineering, Princeton University, Princeton, New Jersey 08544, USA
}
\author{S. A. Lyon}
\affiliation{
    Department of Electrical Engineering, Princeton University, Princeton, New Jersey 08544, USA
}
\date{\today}
\begin{abstract}
    We describe sensitive magnetometry using lumped-element resonators fabricated from a superconducting thin film of
    NbTiN.  Taking advantage of the large kinetic inductance of the superconductor, we demonstrate a continuous
    resonance frequency shift of $27$ MHz for a change in magnetic field of $1.8~\mu$T within a perpendicular background
    field of 60 mT. By using phase-sensitive readout of microwaves transmitted through the sensors, we measure phase
    shifts in real time with a sensitivity of $1$ degree/nT. We present measurements of the noise spectral density of
    the sensors, and find their field sensitivity is at least within one to two orders of magnitude of superconducting
    quantum interference devices operating with zero background field. Our superconducting kinetic inductance
    field-frequency sensors enable real-time magnetometry in the presence of moderate perpendicular background fields up
    to at least 0.2 T. Applications for our sensors include the stabilization of magnetic fields in long coherence
    electron spin resonance measurements and quantum computation.
\end{abstract} 
\maketitle
Disordered superconductors such as NbTiN, TiN and NbN have become ubiquitous in several fields of study due to their
large kinetic inductance and resilience to large background magnetic
fields\cite{annunziata_tunable_2010,samkharadze_high-kinetic-inductance_2016}. Microwave kinetic inductance
detectors\cite{mazin_microwave_2005,szypryt_ultraviolet_2015,zmuidzinas_superconducting_2012} and superconducting
nanowire single-photon detectors\cite{natarajan_superconducting_2012,mccaughanthesis} fabricated from kinetic inductors
are now routinely used in astronomy and imaging. Kinetic inductors can also be used in applications such as
current-sensing\cite{kher_kinetic_2016}, magnetometry\cite{luomahaara_kinetic_2014}, parametric
amplification\cite{ho_eom_wideband_2012,bockstiegel_development_2014}, generation of frequency
combs\cite{erickson_frequency_2014}, and superconducting qubits\cite{hazard_nanowire_2018,peltonen_hybrid_2018}. 

In this work, we take advantage of the kinetic inductance of a thin film of NbTiN to fabricate lumped-element resonators
whose resonance frequencies are strongly dependent on the perpendicular magnetic field, changing by as much as 27 MHz
for a field change of 1.8 $\mu$T. We demonstrate a method for real-time measurement of AC magnetic fields based on
phase-sensitive readout of microwave transmission through the resonators, finding a detection sensitivity of 1
degree/nT. Our Superconducting Kinetic Inductance Field-Frequency Sensors (SKIFFS) are able to operate in perpendicular
background magnetic fields at least as large as 0.2 T, and may find applications in quantum computation, where
superconducting quantum interference devices (SQUIDs) based on Josephson junctions may not be applicable due to the 
large magnetic fields.

\begin{figure}
    \includegraphics[height=1.2in,width=3.37in]{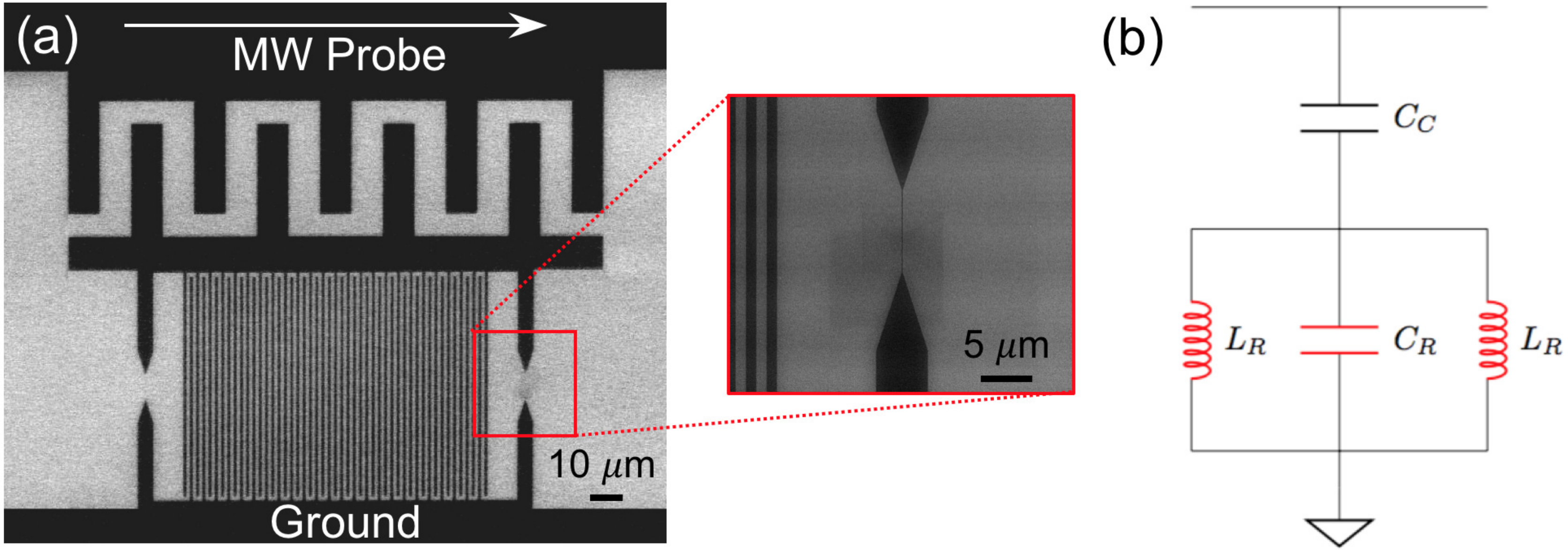}\caption{\label{fig1}(a) Scanning electron
    micrograph of a SKIFFS resonator. Each resonator consists of a 5 \textmicron\ wide
    rectangular superconducting loop that is coupled to a central microwave feedline capacitively using four pairs of
    interdigitated 10 \textmicron\ fingers. The loop inductance is mostly due to the two nanowires that are 100 nm wide
    as shown in the inset. Interdigitated
    capacitors with fingers of width 1 \textmicron\ are placed at the center of the loop to complete the resonator, and
    the lower arm of the superconducting loop is shunted to ground. (b) Lumped-element circuit model of a SKIFFS
    resonator coupled to a microwave feedline via \Cc. The kinetic inductors, \Lr, on each arm of the superconducting
    loop and the capacitor, \Cr, form the resonator.}
\end{figure}

Our SKIFFS are fabricated from a 7 nm NbTiN thin film ($\Tc \sim 9$ K, $R_\text{sheet} =
252~\Omega/\square$) DC-sputtered reactively on a c-axis sapphire substrate using a NbTi alloy target and an Ar/N
environment\cite{starcryo}. The device features are patterned using electron-beam lithography followed by reactive-ion etching
with an SF$_6$/Ar plasma. A scanning electron micrograph of a SKIFFS is shown in figure \figref{fig1}a. The sensor is a
lumped-element microwave resonator fabricated from a 100 \textmicron\ $\times$ 100 \textmicron\ rectangular
superconducting loop defined by a 5 \textmicron\ wide line. Two 100 nm wide nanowires of length $\lnw$ are defined on
the left and right arms of the loop as seen in the inset of \figref{fig1}a. Care is taken to prevent current crowding
at the ends of the nanowires\cite{Hortensius2012,clem_geometry-dependent_2011} by linearly tapering from the micron-wide
loop dimensions to the nanowires over a length of 10 \textmicron. The resonator is completed by placing interdigitated
capacitors at the center of the loop with fingers and gaps of width 1 \textmicron. Each resonator is coupled to a
microwave feedline using a coupling capacitor, which is also an interdigitated structure with four
pairs of 10 \textmicron\ wide fingers that are separated by 10 \textmicron\ gaps.

A lumped-element circuit model of the SKIFFS is shown in \figref{fig1}b. From a Sonnet\cite{sonnet} simulation, we obtain an
estimate of the resonator capacitance, $C_R \approx 0.2$ pF. From the thin-film parameters, we estimate that the
nanowires contribute an inductance of 360 pH/\textmicron, while the micron-sized sections of the loop contribute an
inductance of $7.2$ pH/\textmicron. Using these parameters, we estimate a resonance frequency of $5$ GHz for a sensor
with $\lnw = 10~\mathmicron$.

The device reported in this work has six sensors with varying values of $\lnw$ from 7 to 12 \textmicron\ that are
coupled to a common microwave feedline. Following fabrication, the device is wirebonded to a copper printed circuit
board equipped with microwave connectors and placed in the bore of an external magnet with the surface of the
superconductor perpendicular to the magnetic field, \textBzero. Additionally, a home-made Helmholtz pair coil is
attached to the sample holder such that the generated field, \textBcoil, is parallel to the external field.  We use the
external magnet to generate the moderate background magnetic fields, and the coil to generate small additional magnetic
fields. The magnetic field generated by the home-made coil was calibrated from the shift in the electron spin resonance
line of phosphorus donor electron spins in silicon in a separate experiment.

The device and sample holder assembly are cooled to a temperature of $1.9$ K, and microwave transmission through the
central feedline is monitored using a network analyzer with a microwave power of -72 dBm at the device. All six
resonances are found in the range of $3.945$ to $5.230$ GHz. Microwave powers exceeding -64 dBm were observed to distort
the resonance lineshapes of some of the resonators. This behavior has been observed previously, and is a signature of
the large kinetic inductance of the superconductor\cite{abdo_unusual_2006,swenson_operation_2013}. For the remainder of
this work, we focus on one of the SKIFFS with $\lnw=10~\mathmicron$ and a resonance frequency near 4.2 GHz.

\begin{figure}
    \includegraphics[height=1.65in,width=3.37in]{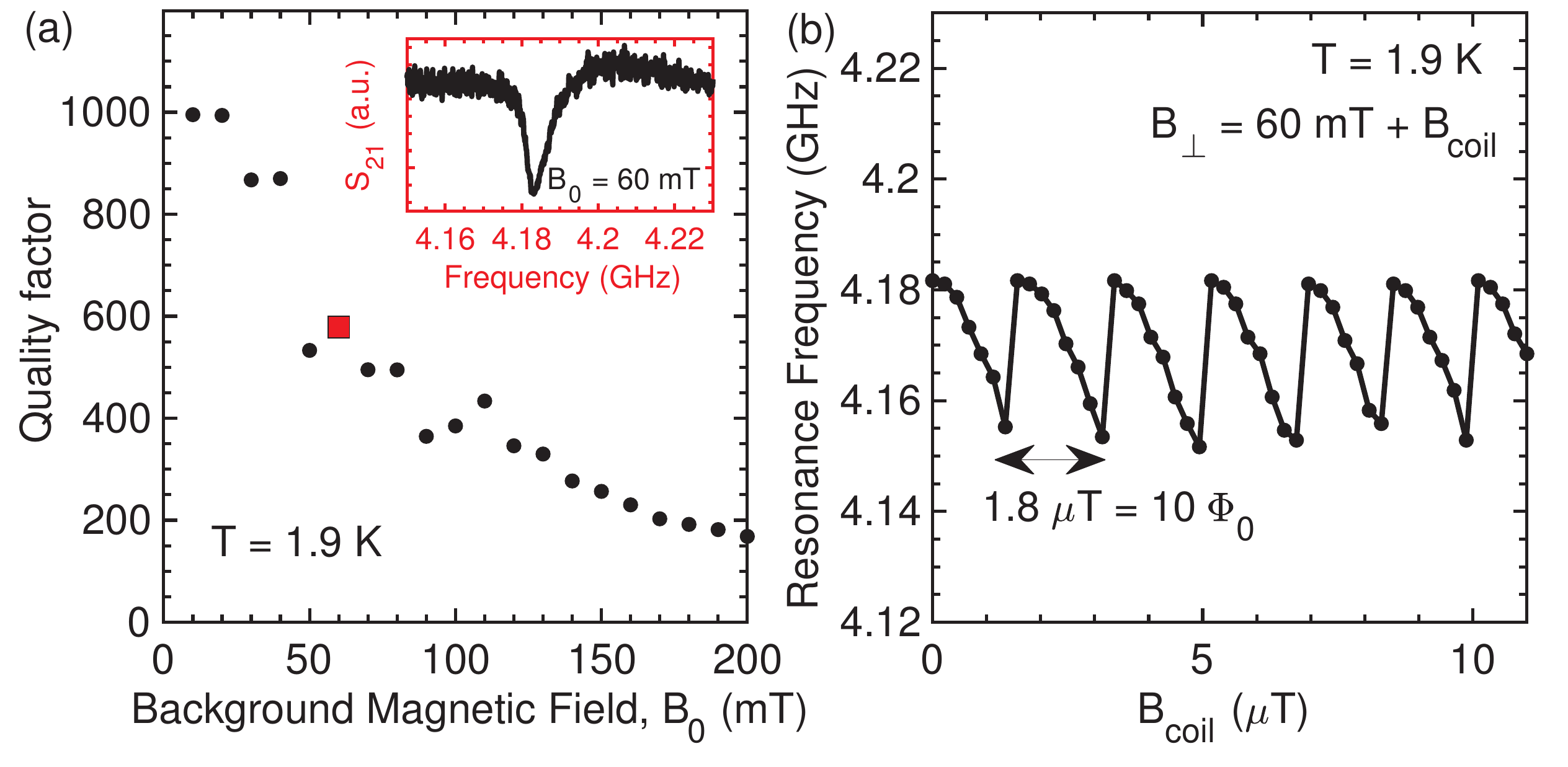}\caption{\label{fig2}(a) Quality factor of a SKIFFS
    resonator as a function of the perpendicular background magnetic field, \textBzero. The inset shows microwave
    transmission through the device near one of the resonance frequencies at $4.182$ GHz for $\mathBzero=60$ mT. (b)
    Shifts in the $4.182$ GHz resonance frequency as a function of small field changes, \textBcoil, applied using a
    home-made Helmholtz pair coil. The perpendicular background magnetic field, \textBzero, is 60 mT. The resonance
    frequency shifts by 27 MHz as \textBcoil\ changes by 1.8 $\mu$T. The abrupt jump in the resonance frequency
    shift results from the finite screening current that can be supported by the superconducting loop at the thin
    nanowire sections, as described in the main text.}
\end{figure}

The loaded quality factor of the resonator depends strongly on the perpendicular magnetic field, \textBzero. In order to
study this dependence, we monitor the resonance near 4.2 GHz as \textBzero\ is swept from 10 to 200 mT. The results are
shown in \figref{fig2}a, where we have extracted the loaded quality factor as a function of the background magnetic
field. As \textBzero\ is increased, the quality factor drops from 1000 at $\mathBzero=10$ mT to 200 at $\mathBzero=200$
mT. The inset shows microwave transmission through the device for $\mathBzero=60$ mT, producing a quality factor 600.
We remark here that the resonator is strongly overcoupled due to the choice of coupling capacitor
parameters. A higher loaded quality factor at zero field can readily be obtained by adjusting the dimensions of the
interdigitated capacitor coupling the resonator to the microwave feedline toward critical coupling. Additionally, higher
quality factors should be possible by operating the resonator at lower temperatures, where $T/\Tc \ll 0.1$. Quality
factors between 10,000 and 100,000 have previously been reported for comparable thin films of NbTiN in the presence of
moderate background fields at 300 mK\cite{samkharadze_high-kinetic-inductance_2016}.

Next, we set $\mathBzero=60$ mT and apply small magnetic fields using our home-made coil. The
results are shown in \figref{fig2}b. As the magnetic field of the coil, \textBcoil, is increased, the resonance
frequency, $f_R$, shifts continuously from its value at 60 mT as
\begin{equation}\label{eq:fVsB}
    \delta f_R(\mathBcoil) = f_{R_0}\left(1-\left(\frac{\mathBcoil}{17~\mu\text{T}}\right)^2\right)
\end{equation}
where $f_{R_0} = 4.182$ GHz is the resonance frequency with $\mathBcoil = 0$ $\mu$T and $\mathBzero=60$ mT. The
functional form of the shift in the resonance frequency has been observed in other
devices\cite{asfaw_multi-frequency_2017,annunziata_tunable_2010,vissers_frequency-tunable_2015,adamyan_tunable_2016,adamyan_tunable_2016,luomahaara_kinetic_2014}.
The resonance frequency shows a maximum shift of 27 MHz as \textBcoil\ is increased to 1.8 $\mu$T. When \textBcoil\
exceeds 1.8 $\mu$T ($\sim$10 flux quanta), the resonance frequency abruptly jumps back to 4.182 GHz and continues to
change as before. This behavior can be understood by noting that a screening current, $\iloop$, is generated in the
superconducting loop in order to keep the magnetic flux threading the loop constant as \textBcoil\ is changed. This
current modulates the kinetic inductance of the superconducting nanowires, \Lr, thereby resulting in a change in the
resonance frequency of the SKIFFS resonator.
Eventually, as \textBcoil\ is increased, $\iloop$ exceeds the critical current of the nanowires. From the
data, we estimate the maximum $\iloop$ to be $\sim 2\mu$A. This results in the formation of a normal metal which breaks
the superconducting loop, allowing additional magnetic flux to thread the loop as the normal section of the loop
returns to the superconducting state.  Once the loop returns to the superconducting state, there is no longer a
screening current since there is no difference between the magnetic flux threading the loop and the applied magnetic
flux due to \textBcoil.

Three observations support this interpretation of the abrupt jumps in $f_R$ as \textBcoil\ is swept. The first is the
functional form in \myeqref{eq:fVsB}, which is similar to the functional form of a kinetic inductor modulated by a DC
current. The screening current, $\iloop$, generated in the superconducting loop is related to the applied magnetic flux,
\textBcoil, by the relation
\begin{equation}\label{eq:devicefluxquantization}
\iloop\LTloop - \mathBcoil A_{\text{loop}} = n\Phi_0
\end{equation}
where $\LTloop$ is the total loop inductance, $A_{\text{loop}}$ is the loop area, $\Phi_0$ is the magnetic flux quantum,
and $n$ is the integer number of flux quanta threading the superconducting loop\cite{luomahaara_kinetic_2014}.
The linear relationship between $\iloop$ and $\mathBcoil$ provides an explanation for the similarity between the
functional form of the change in resonance frequency, $\delta f_R(\mathBcoil)$, to the functional form of a kinetic inductor
that is modulated by a DC
current\cite{asfaw_multi-frequency_2017,annunziata_tunable_2010,vissers_frequency-tunable_2015,adamyan_tunable_2016,adamyan_tunable_2016,luomahaara_kinetic_2014}.
Second, we have observed that the largest value of \textBcoil\ that can be screened before the abrupt jump occurs is
strongly dependent on temperature, and is significantly reduced at higher temperatures. For instance, while $\sim1.8$
$\mu$T
of magnetic field can be screened at a temperature of $1.9$ K with a microwave power of -72 dBm, that value drops to
only $\sim1.2$ $\mu$T at a temperature of $4.2$ K. This observation supports our interpretation since the critical current of
superconductors depends strongly on temperature for temperatures above $\sim0.1\Tc$ \cite{tinkham}.
Third, we have also observed that higher microwave powers reduce the maximum \textBcoil\ that can be supported. This
observation also supports our interpretation since larger microwave powers lead to larger microwave currents, which in turn
limit the largest $\iloop$ that can be supported in the nanowire. We note that similar abrupt
jumps have been observed in the frequency of resonators tuned by nanoSQUIDs fabricated from
Nb constrictions\cite{kennedy_tunable_2018}.

\begin{figure}
    \includegraphics[height=2.6in,width=3.37in]{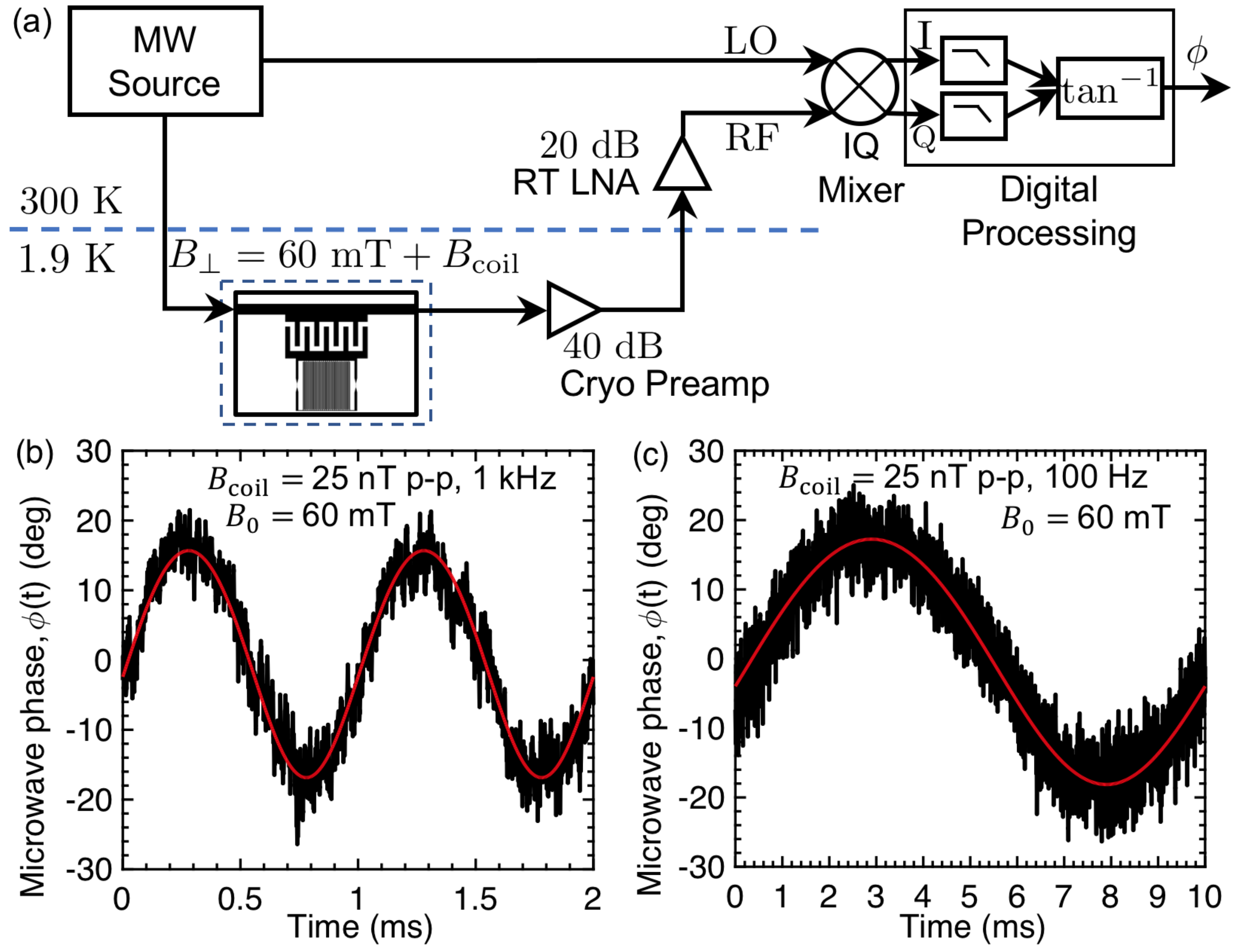}\caption{\label{fig3} (a) Measurement setup for
    real-time detection of small magnetic field fluctuations, $\delta\mathBcoil$, using a SKIFFS resonator. Microwave transmission through the resonator is
    amplified and downconverted using an IQ mixer. The in-phase and quadrature outputs can be used to compute the phase
    of the transmitted microwaves. Changes in the microwave phase transmission are proportional to $\delta\mathBcoil$. Parts (b) and
    (c) show the results of this computation for 25 nT peak-to-peak sinusoidal magnetic fields oscillating at 1
    kHz and 100 Hz, respectively.}
\end{figure}

In order to demonstrate the application of SKIFFS for magnetometry, we use the experimental setup shown in
\figref{fig3}a. A microwave probe tone is applied at the center of the SKIFFS resonance, $f_R$, and transmission through
the device is monitored using phase-sensitive detection. An IQ mixer downconverts the microwave probe tone to DC such
that the phase of microwave transmission through the device can be computed from its in-phase and quadrature components.
For small changes in magnetic field, $\delta\mathBcoil$, the phase of the microwave transmission, $\delta\phi$, follows
linearly with a slope that depends on the quality factor of the resonator. We bias the SKIFFS with $\mathBcoil\approx1.3$
$\mu$T such that the device sensitivity is $\sim31$ MHz/$\mu$T. \figref{fig3}b and c show the results of applying sinusoidal
magnetic fields with amplitude of $\delta\mathBcoil=25$ nT peak-to-peak at a frequency of 1 kHz and 100 Hz,
respectively. The phase of the transmitted microwaves shows oscillations which follow $\delta\mathBcoil$. From the
peak-to-peak change in the phase, we find a phase sensitivity of $\delta\phi/\delta\mathBcoil\approx1$ degree/nT, in
good agreement with an estimate of $0.9$ degrees/nT from the independently measured quality factor of 600 and the bias
point of 31 MHz/$\mu$T.

\begin{figure}
    \includegraphics[height=1.83in,width=3.37in]{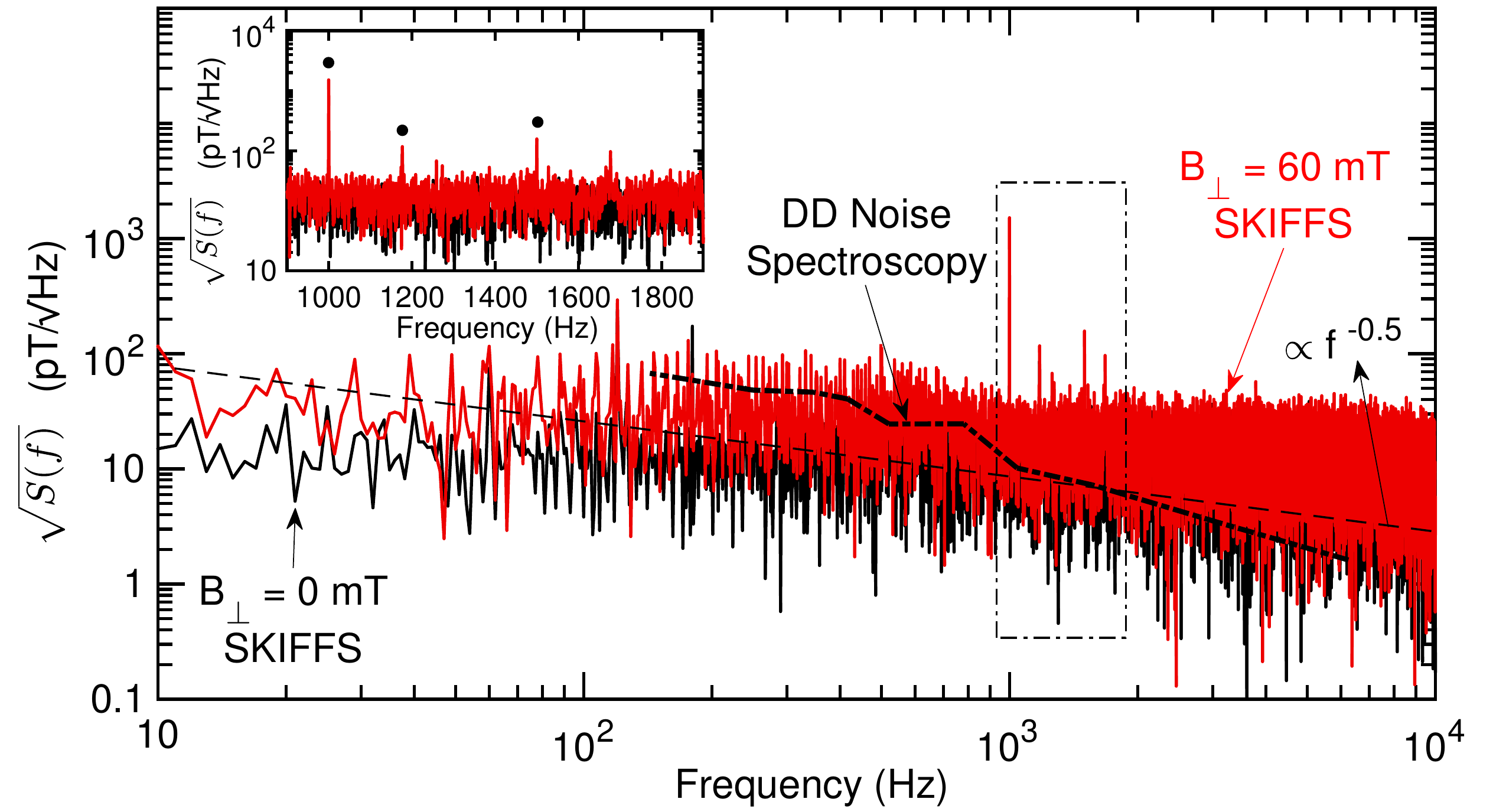}\caption{\label{fig4} Square-root noise power
    spectral density of a SKIFFS resonator, measured by computing the fourier transform of the detected phase signal as
    described in the main text for the cases with and without a background perpendicular magnetic field. The dashed line
    shows a $f^{-0.5}$ fit, indicating that the noise power is of $1/f$-type. The dash-dotted line shows a measurement
    of the fluctuations present in the external magnetic field using dynamical decoupling noise spectroscopy, indicating
    that our measurement of the SKIFFS noise is likely limited by fluctuations in the external magnet. The inset shows a zoom-in near 1 kHz,
    where spurious harmonics of the calibration signal from our waveform generator are apparent, as well as a previously
    characterized source of magnetic field fluctuations in our setup at 1.175 kHz.} 
\end{figure}

We estimate the noise of SKIFFS magnetometry by measuring their phase response to a small 500 Hz sinusoidal magnetic
field. Next, we subtract the sinusoidal component from the measured phase response, and compute the power spectral
density of the remaining noise. The results of this procedure are shown in \figref{fig4} with and without a background
magnetic field, \textBzero, in addition to the sinusoidal excitation. For comparison, we have also included independent
measurements of the magnetic field fluctuations in our experimental setup using dynamical decoupling noise
spectroscopy\cite{biercuk_dynamical_2011,bylander_noise_2011,biercuk_experimental_2009,muhonen_storing_2014}, where the
power spectral density is inferred from the phase accumulated due to magnetic field fluctuations present during an
electron spin resonance experiment with phosphorus donor spins in silicon at 340 mT.

The square-root power spectral density is a factor of 2 larger in the presence of a background magnetic field, likely
due in part to fluctuations in the external magnetic field and vibration-induced noise in our experimental setup.
Measurements at both fields show $f^{-0.5}$ dependence, indicating the presence of 1/f noise power in the detection
similar to low-temperature SQUIDs\cite{clarke2006squid}.  The agreement between the magnetic field noise measured using
SKIFFS and electron spins suggests that our measurement of the SKIFFS noise may be limited by fluctuations in the
external magnetic field\cite{tyryshkin_electron_2003,tyryshkin_coherence_2006}. Further work is necessary to separate
the baseline noise of SKIFFS from the fluctuations in the external magnet. 

The value of the power spectral density in \figref{fig4} is $\sim10$ pT/$\sqrt{\text{Hz}}$
($50~\mu\Phi_0/\sqrt{\text{Hz}}$) at 1 kHz, indicating that our devices are at worst one to two orders of magnitude more
noisy than typical low-temperature SQUIDs at comparable frequencies in zero background field\cite{anton_magnetic_2013}.
Previous work\cite{koch_model_2007, kou_fluxonium-based_2017} has shown that two-level fluctuators can lead to a 1/f noise
power which has a typical value of $S_{\text{B}}^{1/2}(1 \text{ Hz}) = 1~\mu\Phi_0/\sqrt{\text{Hz}}$, indicating the potential for improved sensitivity of our devices by two orders of
magnitude. Kinetic inductance magnetometry with a noise floor of $30$ fT/$\sqrt{\text{Hz}}$ has previously been reported in
a device fabricated from NbN and operated in a shielded environment with zero background field\cite{luomahaara_kinetic_2014}.
The inset in \figref{fig4} shows a close-up of the noise measurement near 1 kHz, where spurious peaks can readily be
identified.  The peaks at 1 and 1.5 kHz are due to the spurious second and third harmonics of the 500 Hz excitation from
our waveform generator. The peak at 1.175 kHz arises from the magnet stabilization system attached to our external
magnet. 

We remark here on practical considerations when using SKIFFS for magnetometry and other applications. 
First, the abrupt jumps shown in \figref{fig2}b limit the dynamic range of the device. The largest magnetic field that
can be applied before an abrupt jump in the resonance frequency is a function of the geometry of the superconducting
loop. The dynamic range can be extended by increasing the width of the nanowires. However, this choice comes at the
expense of a loss in sensitivity, as wider nanowires reduce the kinetic inductance non-linearity of the
loop\cite{asfaw_multi-frequency_2017}. In order to compensate for the reduced sensitivity, the loop area can be
increased. 
Second, the operation of SKIFFS can be extended to even higher background magnetic fields by appropriately designing the
superconducting loop parameters. For our thin-film, we estimate a zero-field London penetration depth of 450 nm. By
choosing parameters for the superconducting loop that are smaller than the penetration depth, the magnetic field
resilience can be extended\cite{samkharadze_high-kinetic-inductance_2016}.  Superconductors with large ($>1$
\textmicron) London penetration depths such as granular aluminum\cite{rotzinger_aluminium-oxide_2017} may extend the use
of SKIFFS to higher magnetic fields. 
Third, we have demonstrated readout of the microwave phase transmission through the SKIFFS devices. The
detection sensitivity of the phase transmission is linearly dependent on the quality factor of the
resonator. As a result, higher detection sensitivity can be achieved by ensuring that the loaded quality factor of the resonators
remains high. For our devices, higher loaded quality factors can be achieved by adjusting the parameters of the coupling
capacitor.
Further improvement should be possible by operating at temperatures $T/\Tc \ll 0.1$, where losses due to thermally
excited quasiparticles and thermally activated motion of trapped flux vortices are significantly reduced.
Finally, we note that even though we have applied small magnetic fields using a
home-made coil, similar results can be obtained by driving DC currents through the central microwave feedline such that
magnetic fields due to the currents couple into the SKIFFS loop. This provides a method of biasing SKIFFS on-chip,
similar to the bias lines used for traditional SQUIDs.

The ability to detect small magnetic fields in the presence of large background fields provides a useful tool for
several areas of research. For example, in the field of quantum computation, previous work has shown that small magnetic
field fluctuations can lead to loss of quantum control of electron spins of donors in silicon as well as trapped ions on
timescales longer than few milliseconds due to undesirable fluctuations in the background magnetic fields that provide
the Zeeman splittings which are necessary for the
experiments\cite{tyryshkin_electron_2003,tyryshkin_coherence_2006,britton_vibration-induced_2016}. In these settings,
SKIFFS can be used to monitor the background magnetic fields and compensate for fluctuations. One approach is to build a
magnetic-field-locked microwave source. In this application, the microwaves transmitted through a SKIFFS resonator can
be used to drive spin rotations by appropriate frequency conversion through a phase-locked loop. More generally, SKIFFS
can be used as resonators that can be tuned using magnetic fields instead of DC currents, and similar structures have
been demonstrated for resonators which are capacitively coupled to a microwave feedline\cite{kennedy_tunable_2018}. 

In summary, we have demonstrated the use of superconducting kinetic inductance field-frequency sensors for magnetometry
in the presence of moderate background magnetic fields. We obtained resonance frequency shifts of $27$ MHz in response to
changes in magnetic field of $1.8~\mu$T. Real-time measurement of AC magnetic fields can be implemented by using
phase-sensitive readout of microwaves transmitted through the sensors with a sensitivity of $1$ degree/nT. We find that
our sensors are at most one to two orders less sensitive in comparison with low-temperature superconducting quantum
interference devices operating in zero background field. We anticipate potential applications of our sensors wherever
small changes in a large background field must be accurately monitored.

See supplementary material for a detailed derivation of \myeqref{eq:devicefluxquantization} and estimates of
the loop inductance and sensitivity of our devices.

We acknowledge helpful discussions with Shyam Shankar. Devices were designed using the CNST nanolithography
toolbox\cite{coimbatore_balram_nanolithography_2016} and fabricated in the Princeton Institute for the Science and
Technology of Materials Micro/Nano Fabrication Laboratory and the Princeton University Quantum Device Nanofabrication
Laboratory. Our work was supported by the NSF, in part through Grant No.\ DMR-1506862, and in part through the Princeton
MRSEC (Grant No. DMR-1420541).

%

\clearpage
\appendix
\begin{center}
\textbf{\large Supplementary Information for \papertitle}
\end{center}
\setcounter{equation}{0}
\setcounter{figure}{0}
\setcounter{table}{0}
\setcounter{page}{1}
\makeatletter
\renewcommand{\theequation}{S\arabic{equation}}
\renewcommand{\thefigure}{S\arabic{figure}}
\renewcommand{\theequation}{S\arabic{equation}}

\subsection{Kinetic Inductance and Flux Quantization}
We begin by noting that the general expression for the kinetic inductance, $L_k$, of a superconducting wire of length $l$ and
cross-sectional area $A$ can be found by equating the total kinetic energy of the Cooper pairs with the standard
expression for the energy stored in an inductor, in complete analogy with magnetic (geometric) inductance \cite{mccaughanthesis}:
\begin{equation}
\frac{1}{2}L_k i^2 = \frac{1}{2}L_k \left(n_s\cdot2e\cdot v_s\cdot A\right)^2 = \frac{1}{2}(2m_e)v_s^2\cdot(n_s\cdot A\cdot l)
\end{equation}
with $m_e$ the mass of an electron, $n_s$ ($v_s$) the density (velocity) of Cooper pairs and $e$ the electronic charge. Solving for $L_k$, we find
\begin{equation}\label{eq:Lk}
L_k = \Lambda\frac{l}{A}
\end{equation}
where $\Lambda = m_e/2n_se^2$.

The supercurrent $i_{\text{loop}}$ flowing through a superconducting loop is related to the magnetic field $\vec{B}$ through the
area enclosed by the loop via the flux quantization condition \cite{tinkham,vanduzer}
\begin{equation}\label{eq:fluxquantization}
\oint_C \Lambda\vec{J}_s \cdot \vec{dl} + \int_s \vec{B}\cdot\vec{ds} = n\Phi_0
\end{equation}
where $J_s$ is the supercurrent density. Here, $\Phi_0$ is the magnetic flux quantum, and the integrals are taken around
the superconducting loop. We note that $n$ can take on integer values.

\subsection{Theory of Device Operation}
For our particular devices, there are four sections within the superconducting loop as described in the main text: two nanowires on either side and
two micron-sized wires completing the loop. 

The first integral in Eq.\ \ref{eq:fluxquantization} can be evaluated by considering the different sections of the
superconducting loop separately. Denoting the loop current as $\iloop$, we write
\begin{equation}
\oint_C \Lambda\vec{J}_s \cdot \vec{dl} = \sum_{\text{sections}} \Lambda\frac{\iloop}{A_\text{section}}\cdot l_{\text{section}}
\end{equation}
where the sum runs over each section of length $l_{\text{section}}$ with cross-sectional area $A_\text{section}$. Using
Eq.\ \ref{eq:Lk}, the first integral in \ref{eq:fluxquantization} reduces to
\begin{equation}\label{eq:p1}
\oint_C \Lambda\vec{J}_s \cdot \vec{dl} = 2\iloop\left(\Lknano + \Lkmicro\right)
\end{equation}
where $\Lknano$ ($\Lkmicro$) is the kinetic inductance of each nanowire (micron-sized) section of the loop. Here, we
have assumed that the two nanowire sections are identical.

Next, we evaluate the second integral in Eq.\ \ref{eq:fluxquantization}. Given the area of the superconducting loop,
$A_{\text{loop}}$, and assuming that the applied field is perpendicular to the loop, 
the total magnetic flux through the loop is the sum of the flux due to the externally applied
magnetic field, $\Phi_{\text{ext}} = -BA_{\text{loop}}$, and the flux due to the screening current generated
by the superconductor in response to the applied field, $\Phi_{\text{resp}} = L_{\text{geom}}i_{\text{loop}}$ where
$L_{\text{geom}}$ is the geometric portion of the total loop inductance. Here, the negative sign indicates that the screening
flux opposes the externally applied flux. Writing $L_\text{geom}$ as the sum of the geometric inductances of the
nanowire and micron-sized sections, $\Lgnano$ and $\Lgmicro$, respectively,
\begin{equation}
L_\text{geom} = 2\left(\Lgnano + \Lgmicro\right) 
\end{equation}
Hence, the second integral in Eq. \ref{eq:fluxquantization} becomes
\begin{equation}\label{eq:p2}
\int_s \vec{B}\cdot\vec{ds} = -BA_{\text{loop}} + 2\iloop\left(\Lgnano + \Lgmicro\right)
\end{equation}
Combining Eqs. \ref{eq:p1} and \ref{eq:p2}, the flux quantization condition in Eq. \ref{eq:fluxquantization} becomes
\begin{equation}
\iloop\LTloop - BA_{\text{loop}} = n\Phi_0
\end{equation}
where $\LTloop = 2\left(\Lknano + \Lgnano + \Lkmicro + \Lgmicro\right)$ is the total loop inductance.

\subsection{Estimating Resonance Frequencies from the Thin Film Parameters}
We estimate the resonance frequency of each sensor as follows. From the sheet resistivity of the thin film
($252~\Omega/\square$), and the critical temperature ($\Tc \sim 9$ K), we estimate a sheet inductance of $L_s = 36
\text{pH}/\square$ and a zero-field London penetration depth of 450
nm\cite{tinkham,zmuidzinas_superconducting_2012,jgaothesis}. From the sheet inductance, we estimate that the nanowire
section of the loop has a kinetic inductance of 360 pH/$\mu$m while the $5~\mu$m-wide section of the loop has a kinetic
inductance of 7.2 pH/$\mu$m.  From the total area of our loop ($\sim 100~\mu\text{m}\times 100~\mu\text{m}$) and for a
$\lnw = 10~\mathmicron$-long section of the nanowire, we estimate the total loop inductance to be $\sim10$ nH.
Simulations using Sonnet software\cite{sonnet} estimate the capacitance of the interdigitated structure at the center of the loop
with $1~\mathmicron$ fingers and gaps to be 0.2 pF, leading to an estimated resonance frequency,
$f_R = \frac{1}{2\pi}\sqrt{\frac{2}{10~\text{nH}\times0.2~\text{pF}}} = 5.03$ GHz. The factor of 2 comes from the parallel
contribution of the left and right arms of the loop to the total loop inductance. The estimated resonance frequency is
in reasonable agreement with the experimentally measured value of 4.182 GHz. A more accurate estimate requires
accounting for the contribution of the 10 \textmicron\ long tapers in the superconducting loop to the total kinetic
inductance.

\subsection{Estimating the Detection Sensitivity of Phase Readout}
Assuming a phase change of $\pi$ radians in $7$ MHz for a resonance centered at $4$ GHz with a quality factor of
600, the slope of the phase transmission through the resonance is $\sim30$ degrees/MHz. In the main text, we have discussed
that the resonators were biased such that the field-frequency conversion factor is $31$ MHz/$\mu$T. Combining these
two values, we estimate the detection sensitivity of phase readout to be $0.9$ degrees/nT. This estimate is in good
agreement with the experimentally observed sensitivity of $1$ degree/nT.

\end{document}